\documentclass[conference]{IEEEtran}
\IEEEoverridecommandlockouts
\usepackage{cite}
\usepackage{amsmath,amssymb,amsfonts}
\usepackage{algorithmic}
\usepackage{algorithm}
\usepackage{graphicx}
\usepackage{textcomp}
\usepackage{xcolor}
\usepackage{booktabs}
\usepackage{url}
\usepackage{hyperref}
\usepackage{multirow}
\usepackage{array}
\usepackage{balance}
\usepackage{listings}
\usepackage{tikz}
\usetikzlibrary{positioning,arrows.meta}
\graphicspath{{./}{figures/}{figures/road/}{figures/representation/}{figures/latency/}{figures/runtime/}}

\newcommand{\eps}{\epsilon}
\def\BibTeX{{\rm B\kern-.05em{\sc i\kern-.025em b}\kern-.08em
    T\kern-.1667em\lower.7ex\hbox{E}\kern-.125emX}}

\begin{document}

\title{Behavioral Residualization for Unsupervised Intrusion Detection in Automotive CAN Networks}

\author{\IEEEauthorblockN{Chandan Hegde}
\IEEEauthorblockA{\textit{Dept. of CSE(AI\&ML)} \\
\textit{Sai Vidya Institute of Technology}\\
Bengaluru, India \\
chandanhegde1007@gmail.com}
\and
\IEEEauthorblockN{Mukundh R Reddy}
\IEEEauthorblockA{\textit{Dept. of CSE(AI\&ML)} \\
\textit{Sai Vidya Institute of Technology}\\
Bengaluru, India \\
mukundhrreddy@gmail.com}
\and
\IEEEauthorblockN{Dr Kumaresh Sheelavant}
\IEEEauthorblockA{\textit{Dept. of CSE(AI\&ML)} \\
\textit{Sai Vidya Institute of Technology}\\
Bengaluru, India \\
kumaresh.s@saividya.ac.in}
}

\maketitle

\begin{abstract}
Modern vehicles exchange safety-critical messages over the Controller Area Network (CAN), a bus designed for low cost and real-time operation that offers neither authentication nor encryption. Once an attacker reaches the network, arbitrary frames can be injected. Existing detectors frequently rely on surface cues (previously unseen identifiers, constant payload bytes, or irregular DLC values). These cues yield near-perfect scores on public corpora whose attacks are separable by exactly those rules, yet they collapse when the adversary simply reuses legitimate identifiers. We introduce per-ID behavioral residualization (PIRD). Fourteen temporal, protocol and payload statistics are extracted from sliding windows belonging to each arbitration ID and then expressed as standardized residuals relative to that ID’s own benign baseline. When the resulting residual vectors are supplied to six standard unsupervised detectors, mean F1 rises in the majority of experimental cells (21 of 24 on HCRL, 30 of 36 on ROAD, five random seeds). On targeted legitimate-ID attacks the method attains recall\,$\geq$\,0.99 and ROC-AUC near 0.99. A deployment-oriented Phase-1 protocol (cross-condition transfer, global threshold, alternative residualization strategies, window/latency sweep) additionally reports Recall@fixed FPR, FPR@target recall, false-alarm rate per hour and detection latency. Two clear failure modes, novel-ID flooding (HCRL DoS, F1\,$=$\,0.02) and cross-ID fuzzing (ROAD, F1\,$=$\,0.27), are quantified and left as open problems.
\end{abstract}


\section{Introduction}
\label{sec:introduction}
Contemporary automobiles contain dozens of Electronic Control Units that exchange messages over the Controller Area Network. Because CAN was standardized primarily for cost and latency (ISO~11898), it provides no cryptographic protection. A compromised diagnostic interface, infotainment system or telematics unit is therefore sufficient to inject arbitrary traffic.

Standards such as UNECE~R155 and ISO/SAE~21434 now treat intrusion detection as an essential complementary control. Many published detectors, however, still depend on presence-oriented signals: previously unseen arbitration IDs, frozen payload bytes or anomalous DLC values. Those signals achieve near-perfect accuracy on the public corpora that happen to contain precisely such artifacts, yet they become unreliable once an attacker re-uses identifiers that already appear in normal traffic.

This paper therefore investigates whether a representation that residualizes each feature against its own arbitration ID’s benign statistics can transfer across both unsupervised detectors and independent datasets that differ in vehicle, collection method and attack realism.

\paragraph{Threat model.}
The adversary is assumed to be a node already attached to the bus (diagnostic port, infotainment or telematics). Following the threat model of~\cite{road}, the adversary may inject frames that carry legitimate identifiers and need not introduce novel IDs, constant bytes or illegal DLC values. Under this assumption, presence-based detectors lose their primary signal and a detector must respond to behavioral deviation.

\paragraph{Contributions.}
\begin{itemize}
    \item A per-ID residual representation in which every feature is expressed relative to the normal moments of its own arbitration ID, so that an injected frame produces a large standardized deviation even when the identifier itself is legitimate. The contribution is the representation; it is evaluated with six interchangeable unsupervised detectors.
    \item Empirical evidence that residualization improves mean F1 in 21 of 24 HCRL cells and 30 of 36 ROAD cells (five seeds) for Isolation Forest, One-Class SVM, LOF, HBOS, Elliptic Envelope and a simple autoencoder.
    \item A per-capture evaluation protocol suited to session-structured corpora such as ROAD.
    \item A quantified robustness boundary: recall\,$\geq$\,0.99 and ROC-AUC\,$\approx$\,0.99 on targeted single-ID attacks that reuse legitimate identifiers, together with explicit failure on novel-ID flooding and cross-ID fuzzing.
    \item A Phase-1 validation suite that measures cross-condition transfer, global thresholding, alternative residualization strategies and the latency/false-alarm trade-off under realistic operating constraints.
\end{itemize}

The remainder of the paper is organized as follows. Section~II surveys related work, Section~III describes the datasets, Section~IV formalizes PIRD, Section~V presents experimental results, and Section~VI concludes.

\section{Related Work}
\label{sec:related}
Existing CAN intrusion detectors can be grouped according to the signal they exploit. Rule- and presence-based approaches look for unusual identifiers, dominant injected IDs, constant payload bytes or illegal DLC values and report near-perfect numbers on HCRL~\cite{hcrl,bari2023}. Those numbers largely reflect the construction of the benchmark; the same detectors lose discriminative power on ROAD~\cite{road}, whose attacks reuse legitimate identifiers. Statistical and one-class methods (One-Class SVM, density estimators) learn a model of normal traffic. PIRD differs by residualizing each feature against the statistics of its own arbitration ID, so that an injected frame on a legitimate ID is no longer masked by that ID’s ordinary variation. Deep sequence models (CANet~\cite{grandad}, contractive autoencoders~\cite{lokman2019}, convolutional~\cite{hcrl}, recurrent, Transformer and graph networks) learn payload structure but are typically trained and tested on a single corpus with labeled attacks, making cross-dataset comparison difficult~\cite{road,lampe2024,rajapaksha2024}. Under an identical protocol a sequence autoencoder does not outperform the lightweight residual detectors on targeted legitimate-ID attacks, indicating that the observed gains originate from the residual features rather than from the choice of Isolation Forest. PIRD is trained exclusively on benign traffic, relies on inexpensive per-ID statistics, and is evaluated on both corpora. To the best of our knowledge no prior study has reported a multi-detector evaluation of per-ID residual features on ROAD under the legitimate-ID-reuse threat model. Recent target-adaptive residual modeling across heterogeneous vehicles~\cite{targetadaptive2026} provides additional support for the residual direction. Foundational analyses of the automotive attack surface~\cite{koscher2010,checkoway2011} motivated subsequent work on entropy~\cite{muter2011}, ID-sequence anomalies~\cite{marchetti2017}, deep classifiers~\cite{kang2016}, LSTM predictors~\cite{taylor2016} and GAN-based detectors~\cite{seo2018}.

\section{Datasets}
\label{sec:datasets}
The two evaluation corpora are summarized in Table~\ref{tab:datasets}.

\begin{table}[t]
\centering
\caption{Summary of the two CAN corpora used for evaluation}
\label{tab:datasets}
\footnotesize
\begin{tabular}{@{}l>{\raggedright\arraybackslash}p{3.0cm}>{\raggedright\arraybackslash}p{3.2cm}@{}}
\toprule
\textbf{Attribute} & \textbf{HCRL Car-Hacking} & \textbf{ROAD} \\
\midrule
Source vehicle 
& Hyundai YF Sonata (scripted) 
& Real vehicle (ORNL dynamometer) \\[2pt]
Message format 
& \texttt{CAN ID DLC data[0..7] label} 
& \texttt{(ts) can0 HEXID\#HEXDATA} \\[2pt]
DLC handling 
& Variable (DLC $<$ 8) 
& Fixed (DLC $=$ 8) \\[2pt]
Attack organization 
& One attack type per file 
& Verified intervals per capture \\[2pt]
Attack categories 
& DoS, Fuzzy, RPM, gear spoof 
& correlated-signal, fuzzing, coolant-temp, speedometer, reverse-light \\[2pt]
Ground-truth source 
& Per-frame class label 
& Interval + ID matching against metadata \\[2pt]
Presence separable? 
& Yes (all four types) 
& No (legitimate AIDs reused) \\
\bottomrule
\end{tabular}
\end{table}

HCRL~\cite{hcrl} was acquired through the OBD-II port and contains four attack types, each with 300 injection intervals of 3--5\,s. ROAD~\cite{road} was recorded on a four-wheel dynamometer at Oak Ridge (approximately 3\,h of ambient traffic across 12 captures and 30\,min of attack traffic across 33 captures). Because ROAD attacks reuse legitimate arbitration IDs, presence-based shortcuts are ineffective. HCRL’s variable DLC requires a DLC-aware parser that places the label at index $3+\text{DLC}$.

\section{Per-ID Behavioral Residual Detection}
\label{sec:method}
\begin{figure}[!t]
\centering
\begin{tikzpicture}[
    >=Stealth,
    font=\sffamily\footnotesize,
    node distance=0.28cm,
    box/.style={
        rectangle, draw=black!65, line width=0.55pt,
        rounded corners=1.5pt, align=center,
        inner sep=4pt, minimum height=0.50cm,
        fill=white, minimum width=3.6cm
    },
    core/.style={
        rectangle, draw=blue!65!black, line width=1.4pt,
        rounded corners=2.3pt, fill=blue!7,
        align=center, inner sep=6.5pt,
        minimum width=6.1cm
    },
    det/.style={
        rectangle, draw=black!50, line width=0.45pt,
        rounded corners=1pt, fill=gray!5,
        align=center, inner sep=2.5pt,
        minimum width=1.50cm, minimum height=0.46cm,
        font=\sffamily\scriptsize
    },
    layer/.style={
        rectangle, draw=black!30, line width=0.55pt, dashed,
        rounded corners=2pt, fill=black!1.2,
        inner sep=5.5pt
    },
    arr/.style={->, line width=0.7pt, draw=black!60},
    sidearr/.style={->, line width=0.65pt, draw=blue!55!black}
]
\node[box] (raw) {Raw CAN Traffic};
\node[box, below=of raw] (stream) {Per-ID Stream Construction};
\node[box, below=of stream] (win) {Sliding Window ($w{=}30$)};
\node[box, below=of win] (feat) {14 Behavioral Features};
\draw[arr] (raw) -- (stream);
\draw[arr] (stream) -- (win);
\draw[arr] (win) -- (feat);
\node[core, below=0.48cm of feat] (core) {
    \textbf{Behavioral Residual Representation (PIRD)}\\[3.5pt]
    {\normalsize $z_{k,j}=\dfrac{x_{k,j}-\mu_{k,j}}{\sigma_{k,j}+\varepsilon}$}\\[2.5pt]
    {\scriptsize Each feature is expressed relative to}\\[-1pt]
    {\scriptsize that ID’s own normal baseline}\\[1.5pt]
    {\scriptsize\bfseries This residual space is the contribution}
};
\draw[arr] (feat) -- (core);
\node[box, draw=blue!50!black, fill=blue!4, minimum width=2.2cm,
      font=\sffamily\scriptsize, left=0.50cm of core] (calib) {
    \textbf{Normal}\\Calibration\\[0.5pt]
    {\scriptsize attack-free}\\[-1pt]
    {\scriptsize per-ID $\mu,\sigma$}
};
\draw[sidearr] (calib.east) -- (core.west);
\node[layer, minimum width=6.7cm, minimum height=1.95cm,
      below=0.52cm of core] (detlayer) {};
\node[font=\sffamily\scriptsize\bfseries, fill=white, inner sep=1.3pt]
    at (detlayer.north) {Any unsupervised detector (interchangeable)};
\node[det] (d1) at ([yshift=0.32cm, xshift=-2.20cm]detlayer.center) {Isolation Forest};
\node[det] (d2) at ([yshift=0.32cm, xshift=0.00cm]detlayer.center)  {One-Class SVM};
\node[det] (d3) at ([yshift=0.32cm, xshift=2.20cm]detlayer.center)  {LOF};
\node[det] (d4) at ([yshift=-0.30cm, xshift=-2.20cm]detlayer.center) {HBOS};
\node[det] (d5) at ([yshift=-0.30cm, xshift=0.00cm]detlayer.center)  {Elliptic Env.};
\node[det] (d6) at ([yshift=-0.30cm, xshift=2.20cm]detlayer.center)  {Autoencoder};
\draw[arr] (core) -- (detlayer.north);
\node[box, minimum width=2.75cm, below=0.42cm of detlayer] (scores) {Anomaly Scores};
\node[box, minimum width=2.75cm, below=0.26cm of scores] (dec) {Attack Decision};
\draw[arr] (detlayer.south) -- (scores);
\draw[arr] (scores) -- (dec);
\end{tikzpicture}
\caption{Conceptual architecture of the proposed framework.}
\label{fig:architecture}
\end{figure}
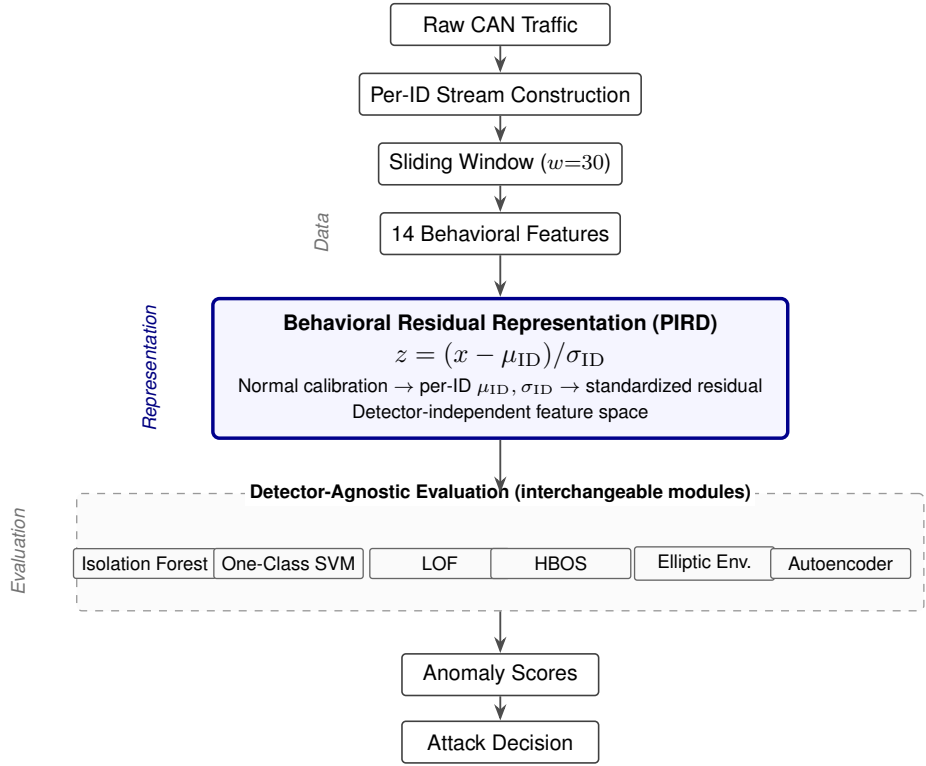

Figure~\ref{fig:architecture} outlines the overall pipeline; the residual representation is the central element. The subsections that follow formalize each stage.

\subsection{Sliding-Window Features}
For every arbitration ID $k$ a deque stores the most recent $w$ frames:
\begin{equation}
    W^{k}_{t}\;=\;\{(\mathbf{p}^{k}_{u},\,t^{k}_{u})\}_{u=t-w+1}^{t}.
\end{equation}
Once the window is full an aggregator $g(\cdot)$ produces the 14-dimensional feature vector
\begin{equation}
    \mathbf{f}^{k}_{t}\;=\; g\big(W^{k}_{t}\big)\;\in\;\mathbb{R}^{14}.
\label{eq:features}
\end{equation}
The features comprise inter-arrival statistics (mean, standard deviation, median, minimum, maximum), DLC statistics (mode, standard deviation), payload statistics (mean, variance, maximum absolute consecutive change, number of distinct byte values, Shannon entropy), the window fill ratio, and the time elapsed since the ID was last observed.

\subsection{Window Label}
A window is marked attack if any of its constituent frames is an attack (``any'' policy). A continuous attack fraction is retained for analysis:
\begin{equation}
    \hat{a}^{k}_{t}=\mathbb{1}\Big[\textstyle\sum_{u} a^{k}_{u}>0\Big],\qquad
    \mathrm{frac}^{k}_{t}=\frac{1}{w}\textstyle\sum_{u} a^{k}_{u}.
\end{equation}

\subsection{Per-ID Residual Transform}
On attack-free calibration windows (at least 20 per ID; otherwise a global fallback is used) the per-feature moments are estimated and every test window is residualized:
\begin{equation}
    r^{k}_{t,j}\;=\;\frac{x^{k}_{t,j}-\mu^{k}_{j}}{\sigma^{k}_{j}+\eps},\qquad
    \mathrm{NaN}\mapsto 0,
\label{eq:residual}
\end{equation}
where $\eps=10^{-6}$ prevents division by zero. When an identifier is absent from the calibration set the global moments $(\mu^{G}_{j},\sigma^{G}_{j})$ are substituted.

\subsection{Detection and Threshold}
An Isolation Forest with 200 trees~\cite{isolation} is trained on the residual vectors. The anomaly score $s(x)=2^{-h(x)/c(n)}$ is negated so that larger values indicate greater anomaly. The decision threshold is the $(1-\mathrm{FPR}_{\mathrm{target}})$-quantile of held-out normal validation scores.

\subsection{Evaluation Protocols}
HCRL is treated as a single continuous stream and split temporally 40/20/40. ROAD consists of independent sessions; residuals are therefore fitted on each capture’s pre-injection normal segment, the detector is trained on additional normal data, and the entire capture is scored. This per-capture protocol is required for fair evaluation of session-structured data.

\section{Experimental Results}
\label{sec:results}

\subsection{HCRL Performance}
Table~\ref{tab:hcrl} reports residual Isolation Forest results at approximately 1\,\% FPR. RPM and gear spoofing are cleanly separated (F1 0.992 and 0.946) because a frozen byte collapses per-window payload entropy. Fuzzy is detected at high recall (0.98) but with elevated FPR (0.15). Novel-ID flooding (DoS) yields essentially no signal (F1 0.02).

\begin{table}[t]
\centering
\caption{Detection performance of residual Isolation Forest on HCRL ($\approx$1\% FPR)}
\label{tab:hcrl}
\footnotesize
\begin{tabular}{lcccccc}
\toprule
\textbf{Attack} & \textbf{F1} & \textbf{Recall} & \textbf{Precision} & \textbf{FPR} & \textbf{ROC-AUC} & \textbf{PR-AUC} \\
\midrule
RPM   & 0.992 & 0.999 & 0.984 & 0.004 & 0.999 & 0.991 \\
gear  & 0.946 & 0.998 & 0.898 & 0.032 & 0.999 & 0.996 \\
Fuzzy & 0.473 & 0.981 & 0.308 & 0.147 & 0.973 & 0.759 \\
DoS   & 0.017 & 0.010 & 0.055 & 0.055 & 0.893 & 0.629 \\
\bottomrule
\end{tabular}
\end{table}

\subsection{Baselines and the Necessity of the Residual Transform}
A naïve presence rule that merely checks for an unseen ID, a dominant ID or a constant byte already exceeds F1 0.99 on every HCRL attack~\cite{hcrl,road}. We therefore treat this figure as an upper-bound presence ceiling; it demonstrates why HCRL alone cannot establish behavioral generalization and motivates evaluation under the stricter legitimate-ID-reuse assumption.

Table~\ref{tab:ablation} shows that raw behavioral features remain near chance while residualization produces the high scores reported above. The residual operator is therefore the decisive component of PIRD.

\begin{table}[t]
\centering
\caption{Effect of residualization: raw features versus residual features (Isolation Forest)}
\label{tab:ablation}
\footnotesize
\begin{tabular}{lccccc}
\toprule
\textbf{Attack} & \textbf{Raw F1} & \textbf{Residual F1} & \textbf{Raw ROC} & \textbf{Residual ROC} & \textbf{Presence ceiling} \\
\midrule
RPM   & 0.000 & 0.992 & 0.855 & 0.999 & $>0.99$ \\
gear  & 0.000 & 0.945 & 0.709 & 0.999 & $>0.99$ \\
Fuzzy & 0.163 & 0.473 & 0.868 & 0.973 & $>0.99$ \\
DoS   & 0.000 & 0.017 & 0.826 & 0.894 & $>0.99$ \\
\bottomrule
\end{tabular}
\end{table}

\subsection{Multi-Detector Representation Validation}
To verify that residualization improves the representation rather than a single algorithm, the identical protocol is repeated across six unsupervised detectors (Table~\ref{tab:multidet}). Residualization improves the large majority of cells; the exceptions are the novel-ID DoS columns and two cells in which residualization slightly lowers an already-strong raw baseline. Gains concentrate on attacks that constitute per-ID deviations and vanish on population-level flooding.

Across five seeds residualization raises mean F1 in 21 of 24 HCRL cells and 30 of 36 ROAD cells (median gains +0.47 and +0.26). A sequence autoencoder evaluated under the same protocol does not surpass the lightweight detectors, confirming that the gains arise from the residual features.

\begin{table}[t]
\centering
\caption{Raw versus residual F1 scores across six unsupervised detectors (HCRL, seed 0)}
\label{tab:multidet}
\footnotesize
\setlength{\tabcolsep}{2.0pt}
\begin{tabular}{lcccccccc}
\toprule
 & \multicolumn{2}{c}{\textbf{DoS}} & \multicolumn{2}{c}{\textbf{Fuzzy}} & \multicolumn{2}{c}{\textbf{RPM}} & \multicolumn{2}{c}{\textbf{gear}} \\
\cmidrule(lr){2-3}\cmidrule(lr){4-5}\cmidrule(lr){6-7}\cmidrule(lr){8-9}
\textbf{Detector} & Raw & Res. & Raw & Res. & Raw & Res. & Raw & Res. \\
\midrule
Isolation Forest  & 0.000 & 0.017 & 0.163 & 0.473 & 0.000 & 0.992 & 0.000 & 0.946 \\
One-Class SVM     & 0.020 & 0.017 & 0.413 & 0.524 & 0.000 & 0.893 & 0.000 & 0.955 \\
LOF               & 0.000 & 0.000 & 0.140 & 0.504 & 0.001 & 0.905 & 0.365 & 0.989 \\
HBOS              & 0.019 & 0.178 & 0.209 & 0.425 & 0.000 & 0.977 & 0.000 & 0.918 \\
Elliptic Envelope & 0.000 & 0.013 & 0.250 & 0.388 & 0.000 & 0.974 & 0.000 & 0.854 \\
Autoencoder       & 0.019 & 0.017 & 0.042 & 0.370 & 0.970 & 0.896 & 0.002 & 0.961 \\
\bottomrule
\end{tabular}
\end{table}

\subsection{Cross-Attack Transfer (Negative Result)}
Table~\ref{tab:cross} shows that only same-type entries are high; the model therefore learns attack-shape-specific norms rather than a universal attack concept.

\begin{table}[t]
\centering
\caption{Cross-attack recall matrix on HCRL (rows = training attack, columns = test attack)}
\label{tab:cross}
\footnotesize
\begin{tabular}{lcccc}
\toprule
\textbf{Train $\backslash$ Test} & \textbf{DoS} & \textbf{Fuzzy} & \textbf{RPM} & \textbf{gear} \\
\midrule
DoS   & 0.010 & 0.235 & 0.000 & 0.000 \\
Fuzzy & 0.010 & 0.981 & 0.000 & 0.000 \\
RPM   & 0.000 & 0.039 & 0.999 & 0.000 \\
gear  & 0.000 & 0.194 & 0.001 & 0.998 \\
\bottomrule
\end{tabular}
\end{table}

\begin{figure}[!htbp]
\centering
\includegraphics[width=0.92\columnwidth]{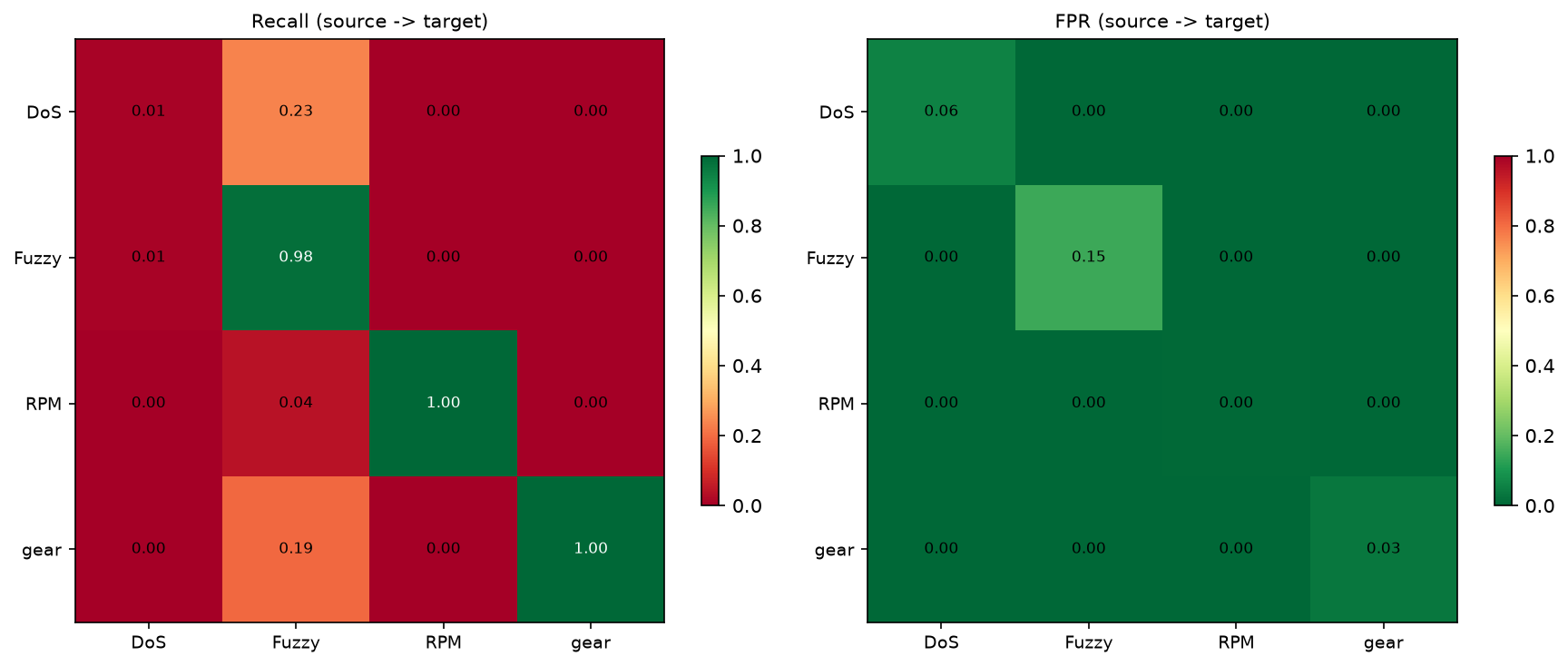}
\caption{Cross-attack recall matrix. Strong diagonal performance and near-zero off-diagonal values confirm that detectors do not transfer across attack types.}
\label{fig:cross_matrix}
\end{figure}

\subsubsection{ROAD results.}
Table~\ref{tab:road} aggregates per-capture residual Isolation Forest performance on ROAD (mean\,$\pm$\,std over the $n$ captures of each attack type).

\begin{table}[t]
\centering
\caption{Per-capture residual Isolation Forest results on ROAD (mean $\pm$ std; single-capture values shown without std)}
\label{tab:road}
\footnotesize
\begin{tabular}{lcccc}
\toprule
\textbf{Attack type} & \textbf{$n$} & \textbf{F1 (mean)} & \textbf{Recall (mean)} & \textbf{FPR (mean)} \\
\midrule
correlated\_signal         & 3 & $0.898\pm0.055$ & $1.000\pm0.000$ & $0.020\pm0.013$ \\
max\_speedometer           & 3 & $0.698\pm0.183$ & $1.000\pm0.001$ & $0.038\pm0.024$ \\
reverse\_light\_off        & 3 & $0.657\pm0.237$ & $1.000\pm0.000$ & $0.048\pm0.032$ \\
reverse\_light\_on         & 3 & $0.544\pm0.124$ & $0.997\pm0.003$ & $0.053\pm0.021$ \\
max\_engine\_coolant\_temp & 1 & $0.265$ & $1.000$ & $0.024$ \\
fuzzing                    & 3 & $0.273\pm0.154$ & $0.340\pm0.333$ & $0.015\pm0.015$ \\
\bottomrule
\end{tabular}
\end{table}

\begin{figure}[!htbp]
\centering
\includegraphics[width=\columnwidth]{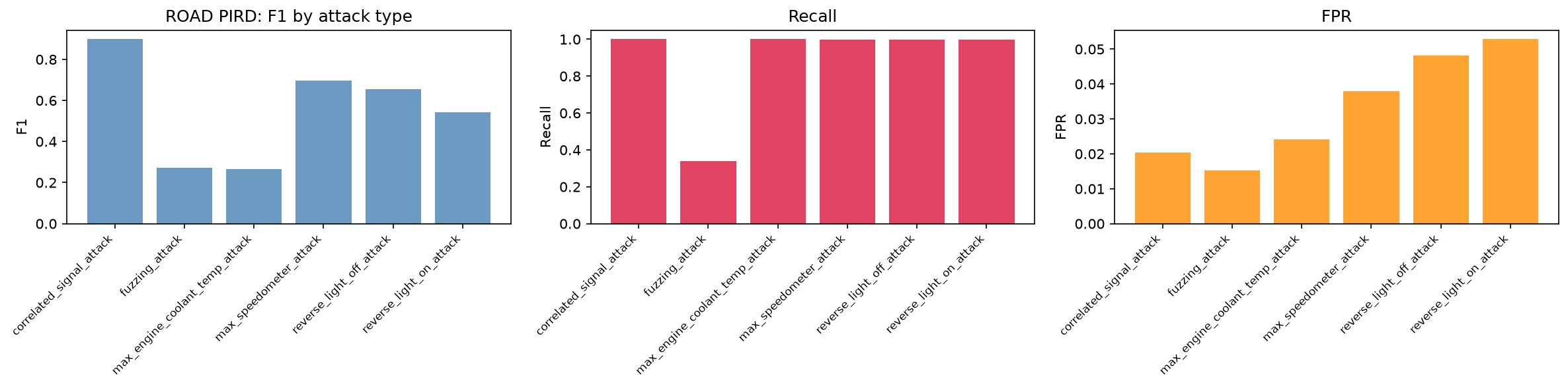}
\caption{PIRD performance on the ROAD dataset by attack type. Targeted single-AID and signal-manipulation attacks achieve near-perfect recall, while cross-ID fuzzing remains challenging by design of the per-ID residual approach.}
\label{fig:road}
\end{figure}

On targeted single-AID attacks recall reaches 1.0 with ROC-AUC near 0.99. The principal remaining bottleneck is residual FPR of 2--5\,\%, which can be reduced by a stricter threshold. Cross-ID fuzzing remains difficult (recall $0.34\pm0.33$) because the residual operator is conditioned per identifier.

\subsection{Efficiency and Coverage Boundary}
Detection latency, throughput and the measured coverage boundary are reported for the residual Isolation Forest at the 1\,\% FPR operating point.

\begin{figure}[!htbp]
\centering
\includegraphics[width=0.95\columnwidth]{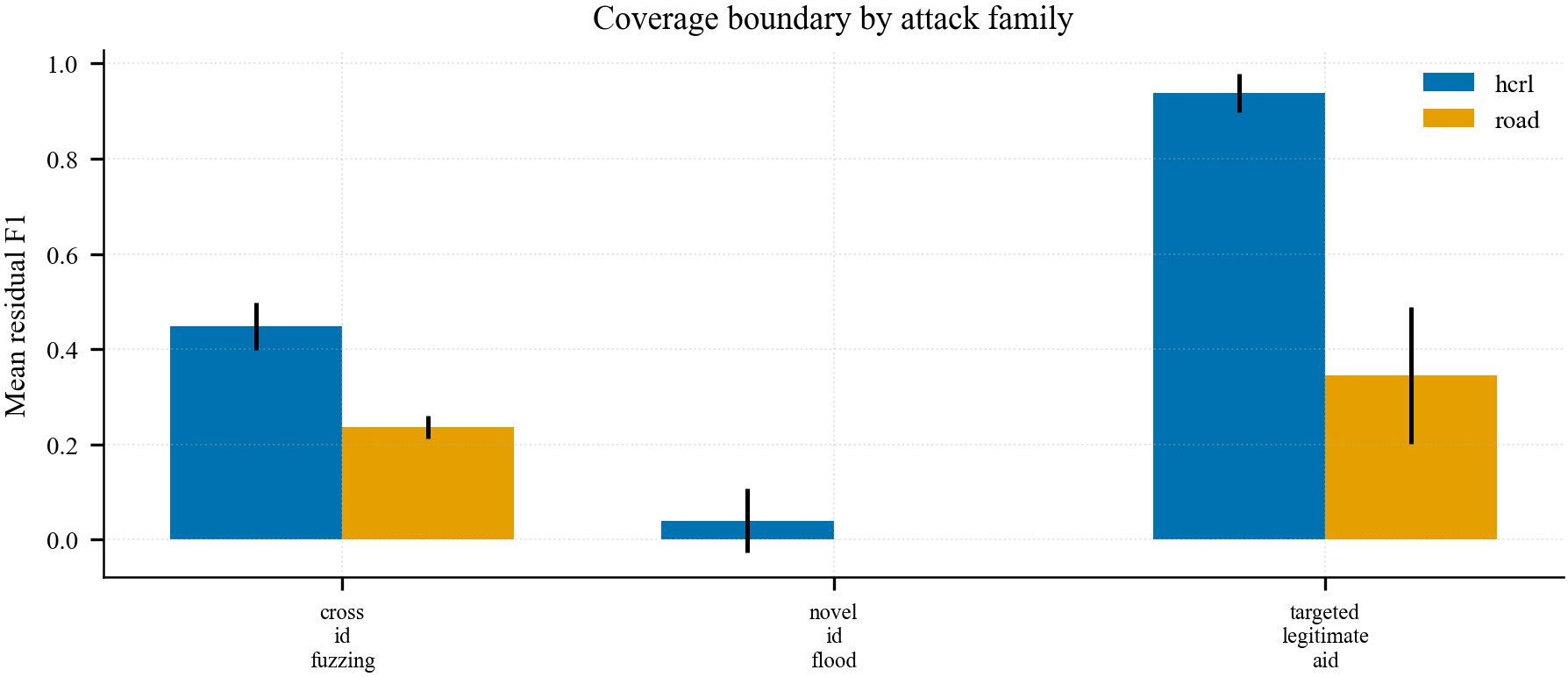}
\caption{Measured coverage boundary of the per-ID residual representation across threat classes and corpora. PIRD is reliable for targeted legitimate-AID attacks, degrades under cross-ID fuzzing, and has no signal for novel-ID flooding.}
\label{fig:coverage_boundary}
\end{figure}

On targeted legitimate-AID attacks PIRD performs strongly on both HCRL (RPM F1 0.992, gear 0.946) and ROAD (ROC-AUC $\geq$0.986 on five of six attack types). Coverage declines on cross-ID fuzzing and collapses on novel-ID flooding.

\begin{figure}[!htbp]
\centering
\includegraphics[width=\columnwidth]{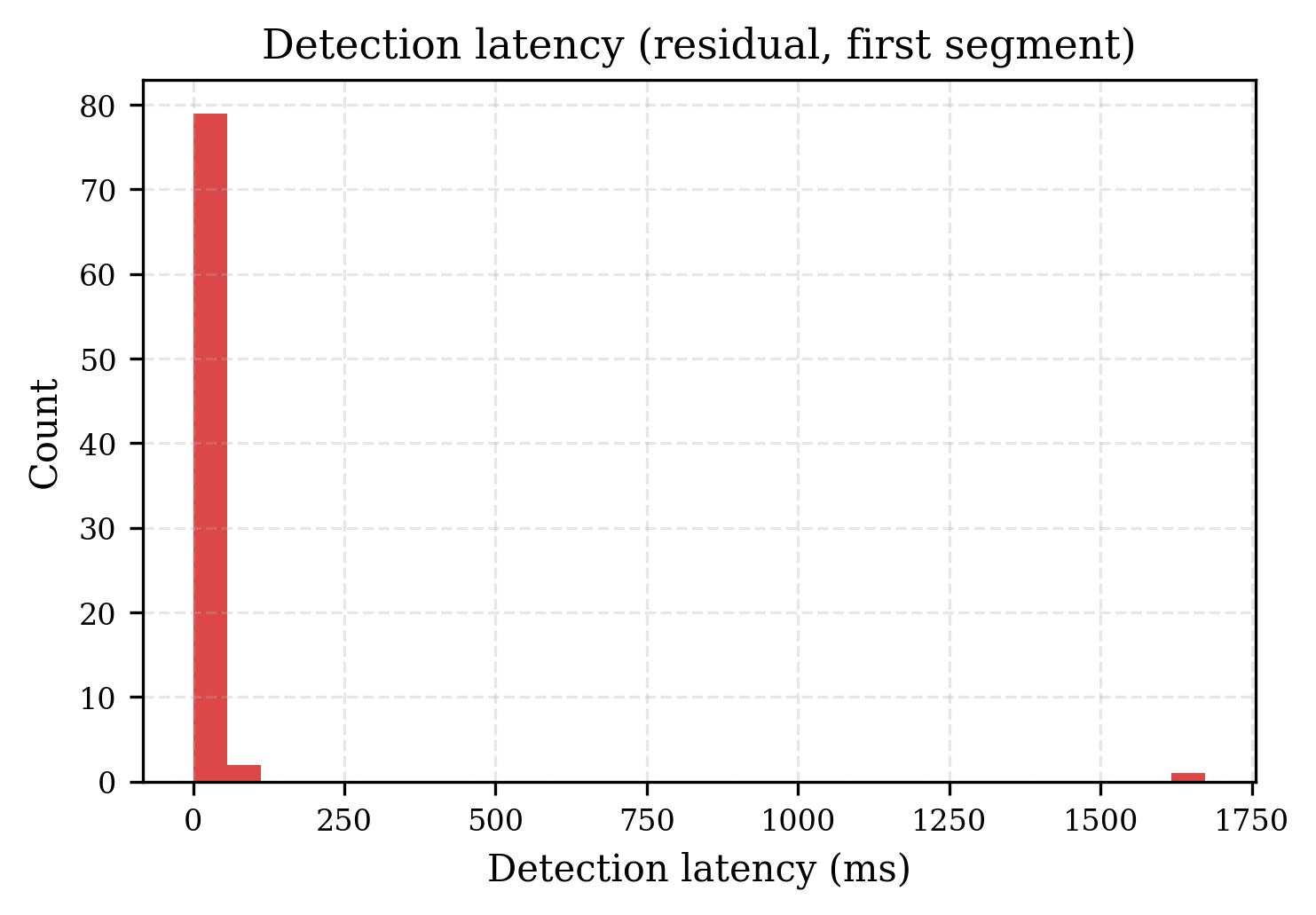}
\caption{Detection-delay histogram (residual Isolation Forest). The distribution is tightly concentrated at zero windows, so an attack is flagged within a single sliding-window step.}
\label{fig:latency_hist}
\end{figure}

Median detection delay on HCRL is zero windows; the worst-case delay is approximately 1.1\,ms.

\begin{figure}[!htbp]
\centering
\includegraphics[width=\columnwidth]{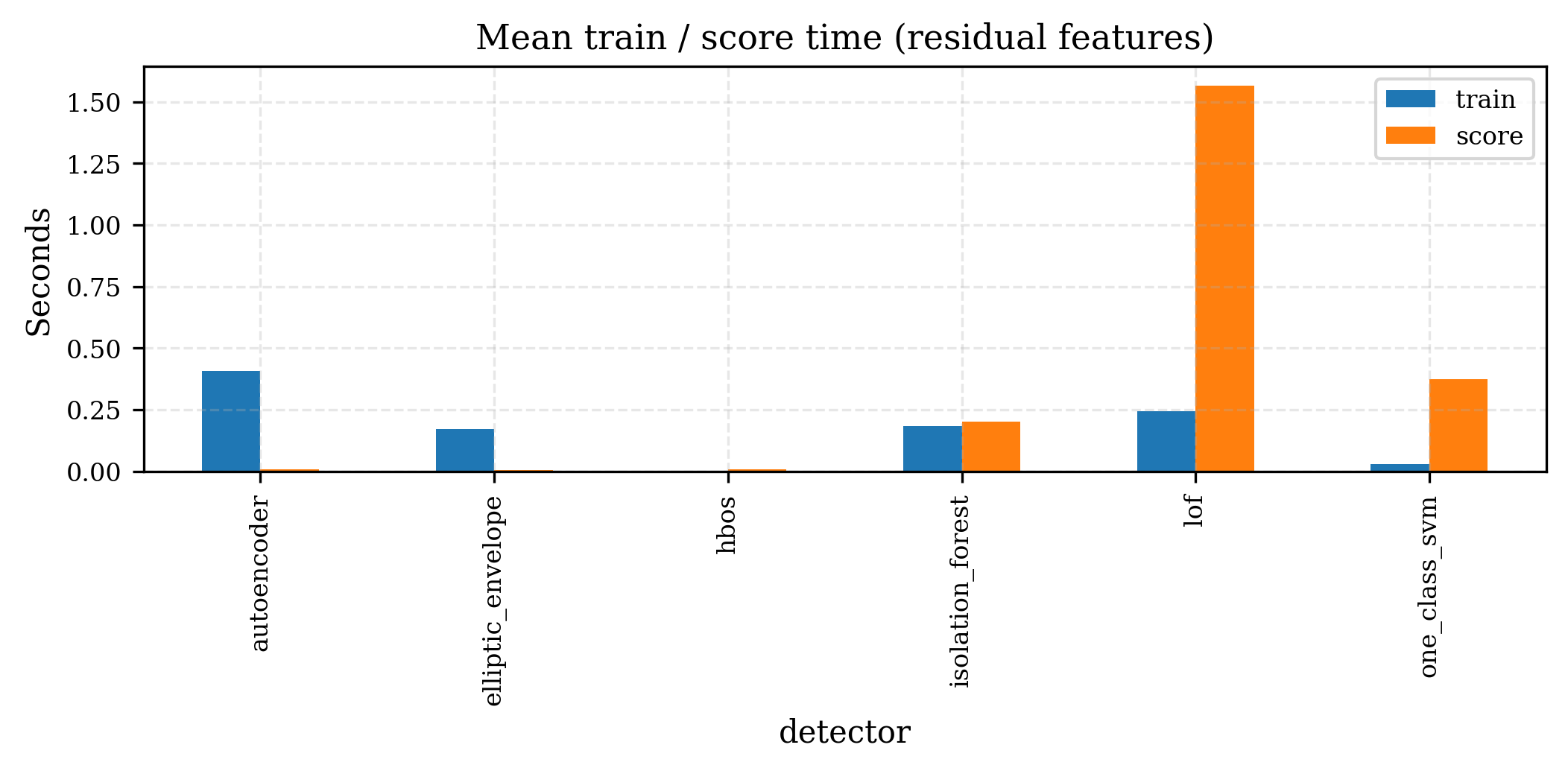}
\caption{Training and scoring time of the Isolation Forest detector on raw versus residual features. The residual transform adds no measurable overhead.}
\label{fig:runtime_train_score}
\end{figure}

Training requires roughly 0.18\,s and scoring of a 20k-window test set requires roughly 0.2\,s ($\approx$0.2\,M windows per second). Peak memory is approximately 525\,MB, dominated by the in-memory window table.

On RPM spoofing the most informative feature is \texttt{iat\_min\_res}; removal under LOFO reduces F1 by 0.236. Features such as \texttt{time\_since\_last\_seen\_res} and \texttt{dlc\_mode\_res} contribute negligibly.

\subsection{Phase-1 Validation: Generalization, Representation, and Deployment}
\label{subsec:phase1}
A Phase-1 protocol examines cross-condition transfer, global thresholding, five residualization strategies and a window/latency sweep, reporting Recall@fixed FPR, FPR@target recall, false alarms per hour and detection latency. Unseen or data-poor identifiers fall back to global statistics, reflecting deployment to a vehicle that was absent from the calibration set.

\subsection{Extended Phases (Compact)}
Beyond Phase 1, three further phases probe the representation. Phase 2 tests adaptive and gradual attacks with a causal global branch (28 features, 200-frame horizon); it recovers population-level attacks (DoS F1 0.50 vs 0.02) but at 24--67\% actual FPR, and never reaches control-adjusted significance under adaptive drift (PIRD $\alpha_{50}=1.17$). Phase 3 evaluates zero-shot transfer across HCRL, ROAD, GEM-CAN and HCRL Survival: transfer fails in every dataset-level direction at a 1\% threshold (target FPR 1.00); ID overlap governs feasibility (Sonata--Kia Soul share 22 IDs, recall 0.94--0.98 but FPR 0.25--0.47; Spark shares 2--3 IDs and collapses to FPR 1.0). Target recalibration restores in-domain F1 to 0.66--0.88 except GEM-CAN (0.15). Phase 4 builds CAN-TFM, an ID-agnostic temporal model (Transformer/GRU/MLP); in-domain F1 is 0.48 and zero-shot mean F1 is 0.144 across 30 source-target cells, too weak to be useful, reported as an early negative result.

\section{Conclusion}
\label{sec:conclusion}
PIRD demonstrates that expressing behavioral features relative to each arbitration ID’s own normal baseline enables effective unsupervised detection of CAN intrusions under the realistic assumption that attackers reuse legitimate identifiers. Ablation isolates the residual operator as the decisive component; the per-capture protocol is required for session-structured data. The approach transfers from the synthetic HCRL corpus to the real ROAD corpus on targeted attacks, while its inability to handle population-level flooding and novelty is reported as a measured coverage boundary. The resulting detector and evaluation methodology constitute a well-characterized foundation for subsequent streaming, multi-vehicle and cross-ID extensions.

\end{document}